\documentclass{cmspaper}

\usepackage{amsmath}
\usepackage{fancybox}
\usepackage{url}

\begin{document}

%
%
%


\begin{titlepage}

   \date{27 July 2009}

  \title{Parametrization of 
the Cosmic Muon Flux for the Generator CMSCGEN}

  \begin{Authlist}
    Philipp Biallass
and Thomas Hebbeker\Instfoot{iiia}{RWTH Aachen University, 
        Physics Institute III A}
  \end{Authlist}

  \begin{abstract}
    The 
    cosmic muon generator CMSCGEN is based on a 
    parametrization of the differential muon flux at ground level, 
    as obtained 
    from the air shower simulation program CORSIKA.
    We present the underlying ansatz for this parameterization
    and provide an approximation of the 
    momentum and angular distributions in terms of 
    simple polynomials, 
    in the momentum range 
    $3 - 3000 \, \mathrm{GeV}$.

  \end{abstract} 
  
\end{titlepage}

\setcounter{page}{2}

\section{Introduction}

The flux $\Phi$ of protons hitting the atmosphere is
steeply falling with energy, approximately $\sim E^{-2.7}$.
This translates into a muon momentum spectrum at the surface
of the earth 
falling roughly as  $p^{-3}$.
On a flat surface the distribution of the zenith angle $\theta$
is to $0^{th}$-order proportional to $|\cos \theta|$, the exact
shape depends on the muon momentum.

The cosmic muon generator CMSCGEN \cite{cmscgen}
was adapted from the fast L3CGEN \cite{l3cgen} program which 
has been written
for the L3 Cosmics project \cite{l3c} some 10 years ago.
These programs are 
based on a parametrization of the cosmic muon flux as a function 
of momentum and zenith angle.
Several developments of the last years suggest to take a fresh
look at this parametrization of the muon flux \cite{l3cgen}:  

\begin{itemize}

\item L3CGEN was successfully used in L3 \cite{l3},
and since 2006 CMSCGEN has been  used\cite{cmscgen}.
Therefore a precise parametrization with known uncertainties
is important, now and in future cosmic ray tests. This is in particular
relevant for collider detectors, for example at the LHC.

\item 
The air shower program CORSIKA \cite{corsika}, which allows the use of
several interaction models, 
has evolved a lot, 
in particular it is now possible to simulate showers with large 
zenith angles, while the old version was limited to muons with an 
angle of at least $30^0$ with respect to the horizon.

\item 
The nuclear interaction models 
available through CORSIKA have been improved
in many ways, and there is a brand-new one, EPOS \cite{epos}, 
which is 
based on recent RHIC data.
 
\item
Precise muon flux measurements \cite{timmermans,l3} 
were made and can be compared to the simulated flux to
assess the precision of 
our parametrization.

\end{itemize}

We study the muon flux for momenta from $3 \, \mathrm{GeV} $
to  $3000 \, \mathrm{GeV} $ at the surface of the LHC ring, 
for all zenith angles.
This note presents and discusses the parametrization obtained
with the CORSIKA version 6.60 \cite{corsika}. 
 
In the following we first 
describe the mathematical form of the muon flux parametrization
and give the values obtained from a CORSIKA 
simulation for primary protons interacting with the atmosphere,
using the default interaction models.
Then we 
compare the results
to experimental data and to simulations based on 
other interaction models and to the parametrization 
of 1998 \cite{l3cgen}.
Finally we discuss the ratio of the  
flux for positive and negative muons and the 
influence of heavier primary nuclei on the momentum spectrum of
the muons. 


\section{Muon Flux Parametrization}

We follow closely the procedure outlined in \cite{l3cgen}:
The total muon flux, integrated over the full range of 
zenith angles (vertical to horizontal) and all azimuthal angles, 
at an 
altitude of about $500 \, \mathrm{m}$ 
is parametrized in the form
\begin{eqnarray}
  \frac{d\, \Phi}{d \, p}
  = C_{norm} \cdot \frac{1}{p^3}  \cdot s(L)
\;\;
. 
\label{integrated_flux}
\end{eqnarray}
$\Phi$ denotes the number of muons per time and area hitting a 
(horizontal) surface area,
$p$ is the muon momentum
and $L = \log_{10} (p/\mathrm{GeV})$.
The 
flux normalization constant
$C_{norm}$ will be discussed later. 
With this parametrization of the steeply falling muon momentum
spectrum the function $s(L)$ (which we need to 
determine)  varies rather little with 
$L$.
The definition (\ref{integrated_flux}) implies that $s(L)$ can be written as
\begin{eqnarray}
  s(L) \sim  p^2 \cdot    \frac{d\, \Phi}{d \, L}
\;\;
.
\end{eqnarray}
This distribution can be obtained from the histogram of 
the logarithm $L$, by applying a weight of 
$(p/\mathrm{GeV})^2 = 10^{2  L}$ to
each muon generated by CORSIKA.
Since the cosmic muon flux falls 
roughly with $p^{-3}$, the expression $s(L)$ is a 
slowly varying function of $L$, and  we can approximate it
by a polynomial:
\begin{eqnarray}
  s(L) = a_0 + a_1 \, L + \ldots + a_6 \, L^6
\;\;\;\;\;\;\;\;\;\;
\end{eqnarray}
It turns out that a polynomial of degree six is sufficient
for our purposes.

The distribution of the azimuthal angle $\phi$ is assumed
to be flat, the dependence on the momentum dependent zenith angle
(here: $\theta= 180^0$ for a vertically downgoing muon)
is parametrized by the normalized function $z(c,L)$ 
with $c = \cos \theta$:
\begin{eqnarray}
  \frac{d\, \Phi}{d c} \sim 
  z(c,L) = b_0(L) + b_1(L) \, c + b_2(L) \, c^2
\;\;\;\;\;\;\;\;\;\;
\label{costheta}
\end{eqnarray}
We use spherical coordinates with the $z$ axis pointing vertically
upwards;
thus vertical muons correspond to $c=-1$, horizontal ones to $c=0$;
these numbers define the range of $c$ values to be considered.
Note that the three 
coefficients $b_i$ are not independent from each other
since the integral 
$ \int  z(c,L) $ 
is one by definition. 
The coefficients $b_i$ 
depend on the muon momentum --- we will discuss this later together
with the Monte Carlo data.

The global constant $C_{norm}$ 
can be derived by comparing the predicted muon flux with
the measured one, for vertical muons at the reference momentum
of $100\, \mathrm{GeV}$, see below.

Thus the full parametrization has the following simple form:
\begin{eqnarray}
\boxed{
  \frac{d\, \Phi}{d  p \, d c \, d \phi}
  = C_{norm} \cdot \frac{1}{p^3}  
\cdot s(L)  \cdot z(c,L)
\cdot \frac{1}{2 \pi}  
}
\label{eq:flux}
\end{eqnarray}


\section{CORSIKA Simulation}

The primary particles impinging on the atmosphere 
are assumed to be protons with an energy distribution $~ E^{-2.7}$.
We use CORSIKA \cite{corsika} with 
the model EPOS (version 1.61) \cite{epos}
for high enery interactions (lab energy
of 80 GeV or more) and GHEISHA (version 2002d) \cite{gheisha} 
for low energy hadronic interactions.
The standard atmosphere is used, zenith angles $\theta$
with $|\cos \theta| > 0.087$ are generated
(corresponding to showers $5^o$ above the horizon). 
In the following we
limit the $\cos \theta$ range to the interval 
$[-1, -0.1]$. 

We have generated five samples of 1 million showers each, 
for different primary energy ranges, covering in total 
all proton energies from $5 \, \mathrm{GeV}$ 
to $10^7 \, \mathrm{GeV}$. Lower or higher values 
hardly contribute to the muon momenta we are investigating 
here \cite{l3cgen}.

\begin{figure}[hbtp]
  \begin{center}
    \resizebox{13cm}{!}{\includegraphics{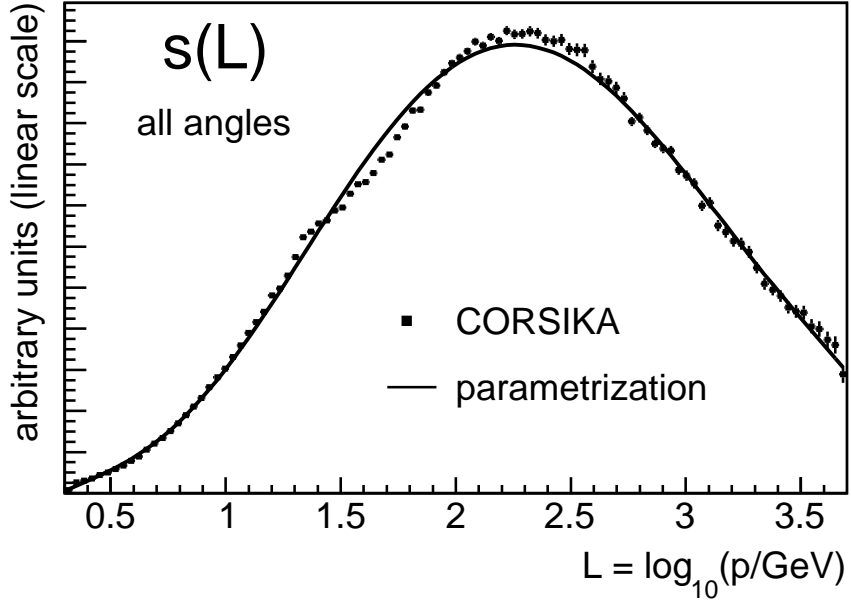}}
    \caption{Momentum spectrum expressed through function $s(L)$,
integrated over zenith angle and azimuth.
The CORSIKA points are obtained with interaction packages EPOS and
GHEISHA. 
}
    \label{fig:spectrum}
  \end{center}
\end{figure}

Figure~\ref{fig:spectrum} 
shows the the function
$s(L)$ as simulated 
for the total muon flux,  
together with the fitted 
polynomial
\begin{eqnarray}
s(L) = -1 
      + 6.2218 \cdot L
      - 13.940 \cdot L^2
      +  18.164 \cdot L^3
      - 9.2278 \cdot L^4
      + 1.9923 \cdot L^5
      - 0.15643 \cdot L^6
\label{s-result}
\end{eqnarray}
Note that we can normalize the coefficients of $s(L)$
arbitrarily; 
here we have done it such that 
the modulus of the first coefficient is one.
%
The fit reproduces the simulated points with
an accuracy of the order of $ 5\%$,
in the
momentum range from $p = 3 \, \mathrm{GeV} \;\; (L \approx 0.5) 
$ to $ p = 3000 \, \mathrm{GeV} \;\; (L \approx 3.5) 
$.
At the 
lowest $L$-bins the fit degrades somewhat (hardly visible in 
Figure~\ref{fig:spectrum}, but the deviations reach up to $20 \%$) 
- we will 
come back to this later. 

\begin{figure}[hbtp]
  \begin{center}
    \resizebox{13cm}{!}{\includegraphics{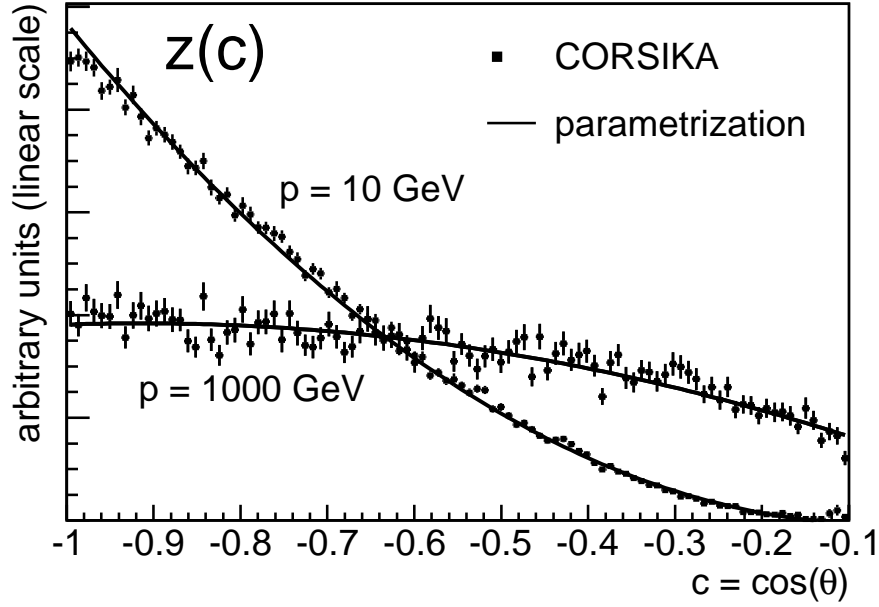}}
    \caption{Distribution $z(c)$ with $c = \cos \theta$ (zenith angle)
for two different muon momenta.
The CORSIKA points are obtained using a combination of the 
interaction packages EPOS and
GHEISHA. 
}
    \label{fig:theta}
  \end{center}
\end{figure}

The $\cos \theta$ distribution $z(c)$ is shown for 
two momentum ranges (near $10 \ \mathrm{GeV}$ and 
around $1000 \ \mathrm{GeV}$) in Figure~\ref{fig:theta}, 
for $ -1 < c < -0.1$. 
Again the CORSIKA points are approximated
quite accurately by the 
parametrization. 
It is evident from Figure \ref{fig:theta} that the momentum
dependence of the zenith angle distribution must be taken into account.
Note that for an isotropic muon flux we would expect 
$z \sim  |\cos \theta|  $, 
due to the {\it flat} 
surface we consider
here (and not a spherical one).

The $L$ dependence of the  coefficients in (\ref{costheta}) 
can again
be parametrized by a polynomial. We obtain:
\begin{eqnarray}
b_0(L) & = & 0.6639  
      - 0.9587 \cdot L
      + 0.2772 \cdot L^2
\nonumber
\\
b_1(L) & = & 5.820 
      - 6.864 \cdot L
      + 1.367 \cdot L^2
\nonumber
\\
b_2(L) & = & 10.39  
      - 8.593 \cdot L
      + 1.547 \cdot L^2
\label{z-result}
\end{eqnarray}
Technical remarks: please 
note that with these fitted curves the normalization requirement
\begin{eqnarray}
   \int_{-1}^{-0.1} \, z(c,L) \, d c   \equiv 1
\end{eqnarray}
is respected only approximately; consequently the $b_i$ 
as given above must 
be renormalized as a function of $L$ before 
$z(c,L)$ can be calculated.
Also, for near horizontal showers and small momenta the flux becomes
very small and the parametrization can result in negative values; this
region in $(L,c)$
should be cut away.

\section{Comparison with Experimental Results}

The flux measurements of the last century
are summarized in the compilation 
\cite{timmermans}. The most precise results
have been obtained
by the L3 collaboration \cite{l3} 
in 2004.

\begin{figure}[hbtp]
  \begin{center}
    \resizebox{13cm}{!}{\includegraphics{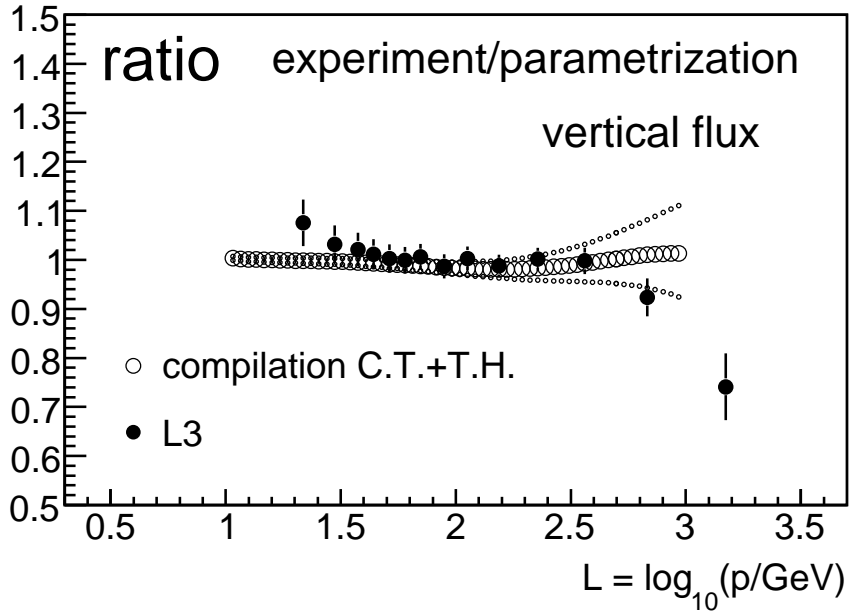}}
    \caption{
Ratio of measured and parametrized vertical muon fluxes.
The small open circles indicate the uncertainties of 
the curve with big open circles, obtained from the compilation 
\cite{timmermans}. The black circles show the results
of the L3 measurements \cite{l3}. 
}
    \label{fig:data_spectrum}
  \end{center}
\end{figure}

In Figure \ref{fig:data_spectrum} we compare the
measured {\it vertical} 
flux\footnote{only the vertical flux has been measured
by several detectors} to our parametrization of $s^v(L)$,
which we have obtained in a similar way as $s(L)$ 
(equation \ref{s-result}) but selecting only (near-)vertical
muons\footnote{
$s^v(L)$ looks similar to $s(L)$ as shown in Figure
\ref{fig:spectrum}, but the peak is shifted slightly to the left.}.
We show the ratio instead of the flux itself 
in order to be more sensitive to potential 
deviations in the distributions.
Note: since we have not yet introduced an absolute
normalization, we have arbitrarily set the ratio to 1 near 
$100 \, \mathrm{GeV} \; (L=2)$.
Since the compilation \cite{timmermans} refers to the flux
at altitude 0 (sea level), we applied the small correction
as given in formula (1) in \cite{timmermans} to 
extrapolate to the altitude of $500 \, \mathrm{m}$. 
Since the L3 
flux is measured at 
a similar altitude, $470 \, \mathrm{m}$, no correction
is necessary.
The agreement between parametrization and measurements is
remarkably good --- only at  momenta above $500 \, \mathrm{GeV}$
deviations 
become visible.

\begin{figure}[hbtp]
  \begin{center}
    \resizebox{13cm}{!}{\includegraphics{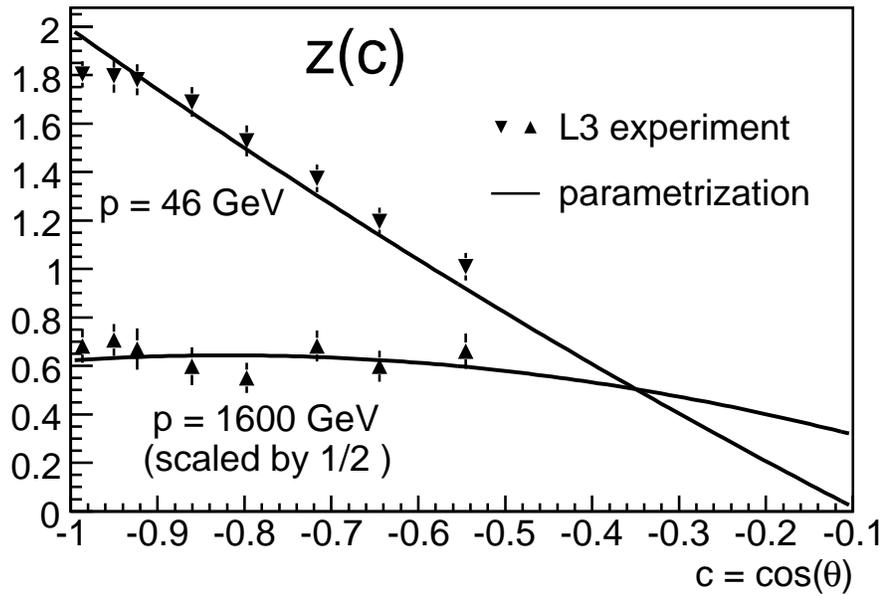}}
    \caption{Distribution $z(c)$ as measured by L3 for
two different momentum values, in comparison with the
parametrization. The $1600 \, \mathrm{GeV}$ 
graph is multiplied by $1/2$ for better visibility.
}
    \label{fig:data_theta}
  \end{center}
\end{figure}

The measured and parametrized $\cos(\theta)$ distributions are compared
in Figure \ref{fig:data_theta}. 
Note that the L3 points have been normalized `by eye'
to the parametrization.
The overall agreement is good, just for (near-)vertical muons
($c = -1\ldots -0.95$) the line is a couple of percent too high
for low momenta.

\section{Comparing Predictions of Different Models}

Finally we compare our parametrization as obtained with EPOS
and GHEISHA to the predictions by other models. 

\begin{figure}[hbtp]
  \begin{center}
    \resizebox{13cm}{!}{\includegraphics{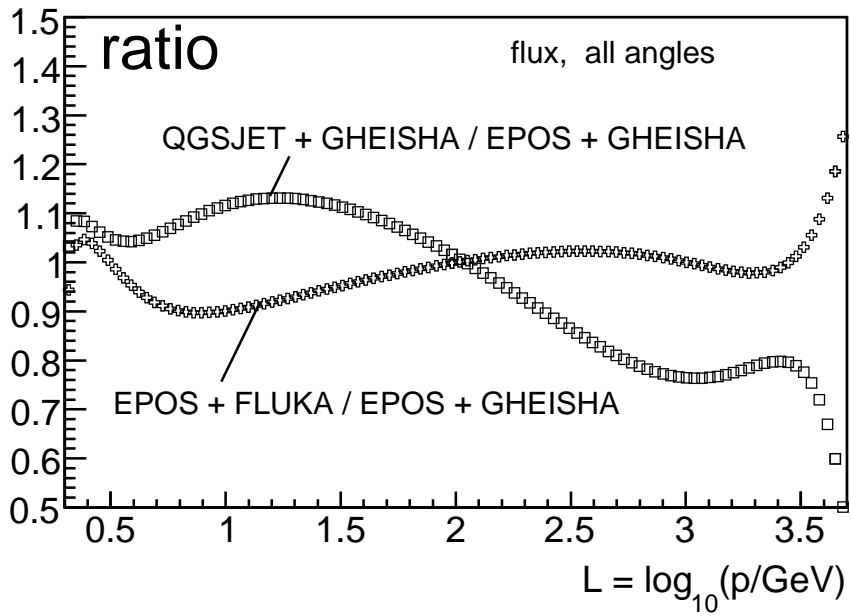}}
    \caption{
Ratio of fluxes for different models within CORSIKA.
}
    \label{fig:spectrum_corsika}
  \end{center}
\end{figure}

First we exchange GHEISHA against the other
low energy hadronic interaction model available in CORSIKA,
FLUKA (version 2006.3b) \cite{fluka}, and we keep EPOS.
In a second simulation run we keep GHEISHA, but exchange 
EPOS against 
QGSJET (version II-03) \cite{qgsjet}.
The result is shown in Figure \ref{fig:spectrum_corsika}
in terms of the ratio of the 
fluxes 
(zenith angle integrated)
predicted by the different models.
Again, we have arbitrarily normalized the ratio to 1 near 
$100 \, \mathrm{GeV} \; (L=2)$.
Differences are clearly visible. The two alternative
models introduced here differ from each other by up to $30\%$
in the momentum range $3 - 3000 \, \mathrm{GeV}$. 
Thus the models are not yet good enough to predict 
the flux with the same precision as
obtained in the recent measurements.
Our reference model, EPOS+GHEISHA, lies 
in 
between the other two models. 
Since the experimental data are in agreement
with the reference model for momenta between $10$ and  
$500 \, \mathrm{GeV}$, 
these 
differences
are only relevant at low and high momenta.
The discrepancies here 
indicate the uncertainties intrinsic in these models.
%

\begin{figure}[hbtp]
  \begin{center}
    \resizebox{13cm}{!}{\includegraphics{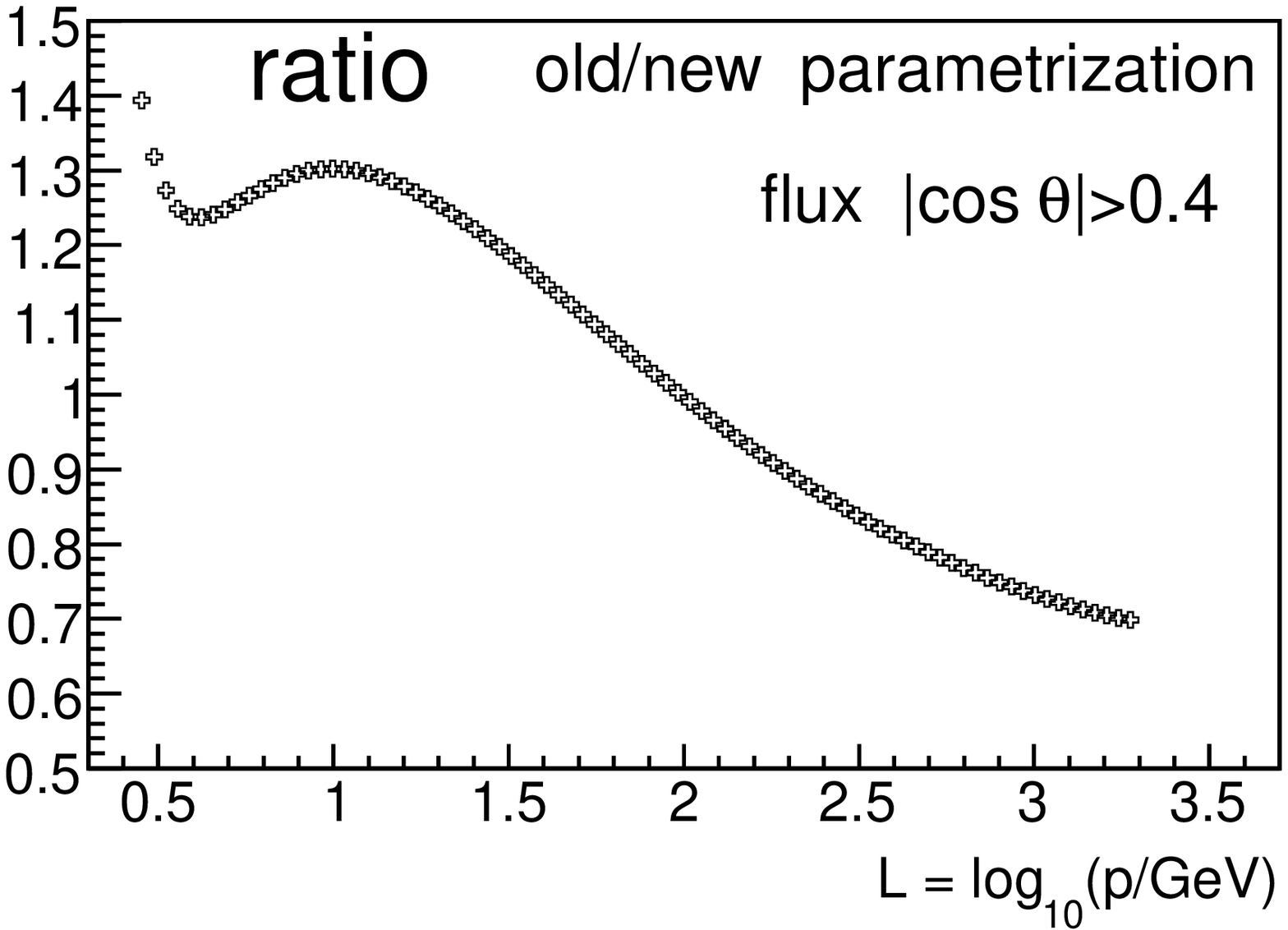}}
    \caption{
Ratio of fluxes for old and new parametrization.
}
    \label{fig:spectrum_old}
  \end{center}
\end{figure}

How does our new parametrization in equation (\ref{s-result})
compare to the old one from 1998 \cite{l3cgen}~?
Figure \ref{fig:spectrum_old}
shows the ratio of the fluxes,
integrated over 
$|\cos \theta| > 0.4$, the zenith angle range used
in the old parametrization.
The differences are quite large,
the new momentum spectrum is significantly harder. Actually,
looking at 
Figures \ref{fig:spectrum_corsika} and \ref{fig:spectrum_old},
there seems to be a general trend: the newer
models (EPOS and FLUKA) give more muons with relatively high
momenta \ldots \ 
Clearly, the discrepancies between the old and new parametrization
make an update of CMSCGEN mandatory.

\begin{figure}[hbtp]
  \begin{center}
    \resizebox{13cm}{!}{\includegraphics{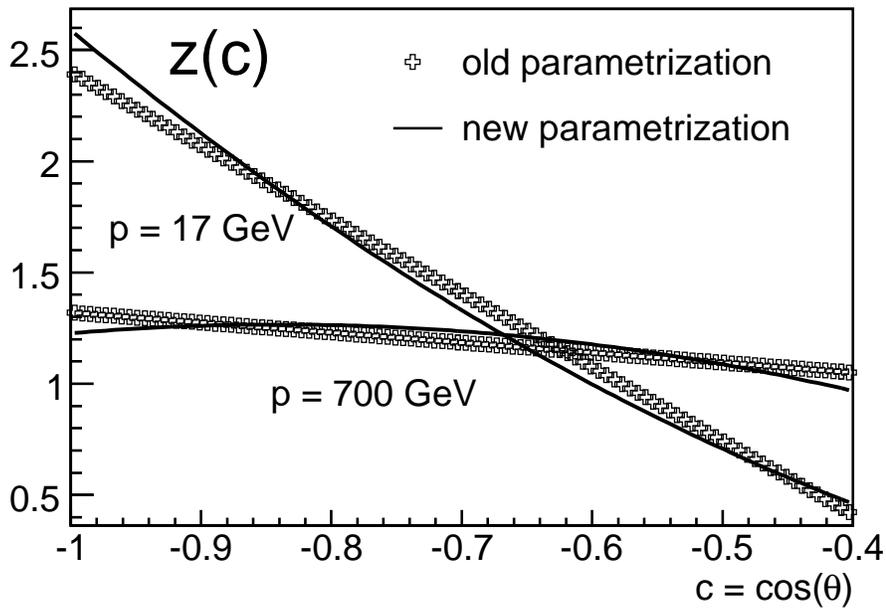}}
    \caption{
$\cos \theta$ distribution --- comparison of new and old
parametrization
}
    \label{fig:theta_old}
  \end{center}
\end{figure}

Figure \ref{fig:theta_old}
compares the parametrizations for the  
$\cos \theta$ distribution.
Shown is the angular range and the momentum values 
as displayed in the
original note \cite{l3cgen}.
The agreement is satisfactory; the small difference in 
shape can be attributed to the fact that before a 
linear approximation was used, while we now use a 
quadratic polynomial.

\section{Muon Flux Normalization}

For muons of momentum $p = 100 \, \mathrm{GeV}$
the vertical flux has been measured 
to \cite{timmermans} 
\begin{eqnarray}
  \frac{d\, \Phi}{d  p \, d \cos\theta \, d \phi}
= (2.59 \pm 0.18) \cdot 10^{-3} 
\, 
\mathrm{m}^{-2} \, \mathrm{s}^{-1} 
\,  \mathrm{GeV}^{-1} \,  \mathrm{sr}^{-1} 
\end{eqnarray}
and \cite{l3}
\begin{eqnarray}
  \frac{d\, \Phi}{d  p \, d \cos\theta \, d \phi}
= (2.63 \pm 0.06) \cdot 10^{-3} 
\,
\mathrm{m}^{-2} \, \mathrm{s}^{-1} 
\,  \mathrm{GeV}^{-1} \,  \mathrm{sr}^{-1} 
\end{eqnarray}
Combining the two results
leads to a flux of vertical muons with the following value, dominated
by the L3 measurement:
\begin{eqnarray}
\boxed{
  \frac{d\, \Phi}{d  p ~ d {\cos\theta} ~ d \phi}
= (2.63 \pm 0.06) \cdot 10^{-3} 
\, 
\mathrm{m}^{-2} \, \mathrm{s}^{-1} 
\,  \mathrm{GeV}^{-1} \,  \mathrm{sr}^{-1} 
}
\label{flux}
\end{eqnarray}

To calculate $C_{norm}$ we first compute the
product
\begin{eqnarray}
P \equiv
\frac{1}{p^3}  
\cdot s(L)  \cdot z(c,L)
\cdot \frac{1}{2 \pi}  
\end{eqnarray}
for $L=2$ and $c=-1$ using our parametrizations
(\ref{s-result}) and (\ref{z-result}):
\begin{eqnarray}
P = 
10^{-6} \, \mathrm{GeV}^{-3}
\cdot 7.094  \cdot 
\left[ -0.1653 - 2.7882 \cdot (-1) - 0.6948 \cdot (-1)^2 \right]
\cdot \frac{1}{2 \pi}
= 2.177  \cdot 10^{-6} \, \mathrm{GeV}^{-3}
\end{eqnarray}
In order to compensate for the small discrepancy at $c=-1$ 
between L3 data and the parametrization of $z(c)$,
which occurs at $46 \, \mathrm{GeV}$, 
see Figure \ref{fig:theta} and Figure \ref{fig:data_theta}, but also 
--- to a lesser extent ---
at $100 \, \mathrm{GeV}$, 
we introduce a `fudge factor'
of 1.05 to optimize the overall agreement of the 
parametrization with the L3 results:
\begin{eqnarray}
P  \, \to \, P / 1.05 = 2.073 \cdot 10^{-6} \, \mathrm{GeV}^{-3}
\end{eqnarray}
 Comparing with (\ref{eq:flux}) and (\ref{flux}) gives:
\begin{eqnarray}
C_{norm}
= 1.27 \cdot 10^{3} 
\,
\mathrm{GeV}^{2} \, \mathrm{m}^{-2} 
\,  \mathrm{s}^{-1} \,  \mathrm{sr}^{-1} 
\end{eqnarray}
 
\section{Charge Ratio}

The measurements yield a ratio $R$ of positive and negative
muon fluxes of \cite{timmermans} 
\begin{eqnarray}
 R = 1.268 \pm 0.028
\end{eqnarray}
and \cite{l3}
\begin{eqnarray}
 R = 1.285 \pm 0.019
\end{eqnarray}
resulting in a combined value of
\begin{eqnarray}
\boxed{ 
R = 1.280 \pm 0.016
}
\end{eqnarray}
for vertical incidence and momenta around $100 \, \mathrm{GeV}$.
The data are consistent with a charge ratio $R$ which is
independent of momentum (at least from 
$10$ to $500 \, \mathrm{GeV}$) 
and also of zenith angle, if one 
stays away from near horizontal muons \cite{l3,grieder}.

The CORSIKA simulations (EPOS+GHEISHA) yield R values
varying by $\pm 5\% $ 
around 1.40 for momenta  
in the range 
$10$ to $1000 \, \mathrm{GeV}$, 
in disagreement with
the measurements.

For the generator CMSCGEN the experimental 
result of $R = 1.28$ must be used, i.e. the total muon flux
must be split into positive and negative muons with the relative
fractions of 
\begin{eqnarray}
  f^+ = 66.1 \, \%
\;\;\;\;\;\;\;\;
  f^- = 43.9 \, \%
\;\; 
.
\end{eqnarray}

\section{Heavy Nuclei}

It was already shown in \cite{l3cgen} that primary helium or iron
nuclei yield muon spectra that are similar to those from protons.
Since anyway the parametrization presented here describes the
experimental data well, there was no need to repeat such a study here.


\section{Discussion of Uncertainties}

Our parametrizations for $z(c)$ can be used from $c = -1$
(vertically downward going muons)
to $c = -0.1$. For higher values (nearly horizontal showers)
the Jura mountains introduce a $\phi$ dependent 
absorption of muons, since they stand up about $10^o$
above the horizon, seen e.g. from the CMS\cite{cms} site.  

For momenta below $3 \, \mathrm{GeV} \; (L \approx 0.5)$ 
and above $3000 \, \mathrm{GeV} \; (L \approx 3.5)$ 
our parametrization is not valid and should therefore 
{\bf not} be used.
Reasons are the instability of the fit for low and high momenta
(see Figures \ref{fig:spectrum_corsika} and \ref{fig:spectrum_old}) 
and the lack of 
reliable experimental data in these regimes (see Figure \ref{fig:data_spectrum}).

In the `central' momentum range from
$10$ to $500 \, \mathrm{GeV}$ our parametrization seems to
work quite well: 
the measured cosmic fluxes are reproduced 
at the 
$5- 10\%$ level. 
So we 
assign 
an uncertainty ($68\%$ confidence level) 
to the absolute differential flux as given by our parametrization
of 
$ \pm 7 \%$.

In the `fringe' momentum regions 
(below $10 \, \mathrm{GeV}$ and above $500 \, \mathrm{GeV}$)
the uncertainty increases rapidly.
Due to the missing
experimental support at high momenta we assign a relative 
uncertainty to the flux increasing up to $\pm 50 \% $ at
$3000 \, \mathrm{GeV}$. This
estimate is obtained by extrapolating
the uncertainties displayed in Figure \ref{fig:data_spectrum}.
At low momenta geomagnetic effects become important \cite{grieder} and atmospheric
and solar influences 
make the muon flux vary with time \cite{timmermans}, 
there is no comparison to experiment either, and the model
predictions disagree with each other, see above.
Therefore we attribute
to the
low momentum regime 
a relative uncertainty 
increasing from $\pm 7\% $ at $10 \, \mathrm{GeV}$ 
up to $\pm 25\% $ for 
$p = 3 \, \mathrm{GeV}$.



\section{Acknowledgements}

We thank Dieter Heck, Tanguy Pierog, Ralf Ulrich and Hans
Dembinski for technical help with the CORSIKA program.


\end{document}